%% file: distrib.tex
\documentclass[aps,showpacs,showkeys,superscriptaddress,11pt]{revtex4}

\usepackage{graphicx,epic,eepic,epsfig,amsmath,latexsym,amssymb,verbatim}

\usepackage{theorem}
\newtheorem{definition}{Definition}[section]
\newtheorem{proposition}[definition]{Proposition}
\newtheorem{lemma}[definition]{Lemma}

\newtheorem{theorem}[definition]{Theorem}

\def\squareforqed{\hbox{\rlap{$\sqcap$}$\sqcup$}}
\def\qed{\ifmmode\squareforqed\else{\unskip\nobreak\hfil
\penalty50\hskip1em\null\nobreak\hfil\squareforqed
\parfillskip=0pt\finalhyphendemerits=0\endgraf}\fi}
\def\endenv{\ifmmode\;\else{\unskip\nobreak\hfil
\penalty50\hskip1em\null\nobreak\hfil\;
\parfillskip=0pt\finalhyphendemerits=0\endgraf}\fi}
\newenvironment{proof}{\noindent \textbf{{Proof~} }}{\qed}

\newenvironment{example}{\smallskip \noindent \textbf{{Example~}}}{}

\newcommand{\exampleTitle}[1]{\textbf{(#1)}}

\mathchardef\ordinarycolon\mathcode`\:
\mathcode`\:=\string"8000
\def\vcentcolon{\mathrel{\mathop\ordinarycolon}}
\begingroup \catcode`\:=\active
  \lowercase{\endgroup
  \let :\vcentcolon
  }

\newcommand{\nc}{\newcommand}
\nc{\rnc}{\renewcommand}
\nc{\beq}{\begin{equation}}
\nc{\eeq}{{\end{equation}}}
\nc{\beqa}{\begin{eqnarray}}
\nc{\eeqa}{\end{eqnarray}}
\nc{\lbar}[1]{\overline{#1}}
\nc{\bra}[1]{\langle#1|}
\nc{\ket}[1]{|#1\rangle}
\nc{\ketbra}[2]{|#1\rangle\!\langle#2|}
\nc{\braket}[2]{\langle#1|#2\rangle}
\nc{\proj}[1]{| #1\rangle\!\langle #1 |}
\nc{\avg}[1]{\langle#1\rangle}
\rnc{\max}{\operatorname{max}}
\nc{\Rank}{\operatorname{Rank}}
\nc{\smfrac}[2]{\mbox{$\frac{#1}{#2}$}}
\nc{\Tr}{\operatorname{Tr}}
\nc{\ox}{\otimes}
\nc{\dg}{\dagger}
\nc{\dn}{\downarrow}
\nc{\cA}{{\cal A}}
\nc{\cB}{{\cal B}}
\nc{\cC}{{\cal C}}
\nc{\cD}{{\cal D}}
\nc{\cE}{{\cal E}}
\nc{\cF}{{\cal F}}
\nc{\cG}{{\cal G}}
\nc{\cH}{{\cal H}}
\nc{\cI}{{\cal I}}
\nc{\cJ}{{\cal J}}
\nc{\cK}{{\cal K}}
\nc{\cL}{{\cal L}}
\nc{\cM}{{\cal M}}
\nc{\cO}{{\cal O}}
\nc{\cP}{{\cal P}}
\nc{\cR}{{\cal R}}
\nc{\cS}{{\cal S}}
\nc{\cT}{{\cal T}}
\nc{\cX}{{\cal X}}
\nc{\cZ}{{\cal Z}}
\nc{\optimal}{^*}
\nc{\csupp}{{\operatorname{csupp}}}
\nc{\qsupp}{{\operatorname{qsupp}}}
\nc{\esupp}{{\operatorname{esupp}}}
\nc{\var}{\operatorname{var}}
\nc{\rar}{\rightarrow}
\nc{\lrar}{\longrightarrow}
\nc{\Span}{\operatorname{Span}}
\rnc{\vec}{}

\def\a{\alpha}
\def\b{\beta}

\def\d{\delta}
\def\e{\epsilon}

\def\l{\lambda}

\def\x{\xi}

\def\r{\rho}
\def\s{\sigma}

\def\ph{\varphi}

\def\ps{\psi}
\def\o{\omega}

\nc{\RR}{{{\mathbb R}}}
\nc{\CC}{{{\mathbb C}}}
\nc{\FF}{{{\mathbb F}}}
\nc{\NN}{{{\mathbb N}}}
\nc{\ZZ}{{{\mathbb Z}}}
\nc{\PP}{{{\mathbb P}}}
\nc{\QQ}{{{\mathbb Q}}}
\nc{\UU}{{{\mathbb U}}}
\nc{\EE}{{{\mathbb E}}}
\nc{\id}{{\mathbb I}}

\begin{document}

\title{{\Large On the distributed compression of quantum information}}

\author{Charlene Ahn}
\email{cahn@theory.caltech.edu}
\affiliation{Institute for Quantum Information, Caltech 107-81, Pasadena, CA 91125, USA}
\author{Andrew Doherty}
\email{dohertya@caltech.edu}
\affiliation{Institute for Quantum Information, Caltech 107-81, Pasadena, CA 91125, USA}
\affiliation{School of Physical Sciences, University
  of Queensland, Brisbane 4072, Australia}
\author{Patrick Hayden}
\email{patrick@cs.mcgill.ca} \affiliation{Institute for Quantum
Information, Caltech 107-81, Pasadena, CA 91125, USA}
\affiliation{School of Computer Science, McGill University,
Montreal, Canada}
\author{Andreas Winter}
\email{a.j.winter@bris.ac.uk}
\affiliation{Department of Mathematics, University of Bristol,
  University Walk, Bristol BS8 1TW, U. K.}

\date{September 18, 2005}

\begin{abstract}
We consider the problem of distributed compression for correlated
quantum sources. The classical version of this problem was solved by
Slepian and Wolf, who showed that distributed compression could take
full advantage of redundancy in the local sources created by the
presence of correlations. We show that, in general, this is not the
case for quantum sources by proving a lower bound on the rate sum
for irreducible sources of product states which is stronger than the
one given by a naive application of Slepian-Wolf. Nonetheless,
strategies taking advantage of correlation do exist for some special
classes of quantum sources. For example, Devetak and Winter
demonstrated the existence of such a strategy when one of the
sources is classical. Here we find optimal non-trivial strategies
for a different extreme, sources of Bell states. In addition, we
illustrate how distributed compression is connected to other
problems in quantum information theory, including
information-disturbance questions, entanglement distillation and
quantum error correction.
\end{abstract}



\maketitle


\section{Introduction} \label{sec:intro}

The insights that have come from efforts to study quantum mechanics
from an information-theoretic point of view are profound and wide-ranging,
demonstrating that quantum information can be compressed~\cite{Schumacher95,JozsaS94},
stabilised~\cite{CalderbankS96} and usefully processed~\cite{Shor94}.
Schumacher's theorem~\cite{OhyaP93,Schumacher95,JozsaS94}, in particular,
demonstrated the fungibility of quantum states by quantifying their
compressibility, justifying the use of the \emph{qubit} as the fundamental
unit of quantum information.

In this paper we consider a distributed variant of the problem
posed by Schumacher. Namely, we suppose that a source distributes
quantum states to two or more parties, who independently compress
the states before sending them on to a receiver, who is required
to be able to reconstruct the original inputs. Since many ideas
for the design of quantum computers and other quantum information
processing devices envision a network of
relatively small quantum processors sending quantum information
between nodes~\cite{Cirac97,SasuraB01}, finding good
protocols for distributed compression of quantum data could
conceivably have important practical benefits. More generally,
much of quantum information theory is concerned with the
manipulation of data under locality
constraints~\cite{BennettDSW96}, so our problem connects naturally
to these investigations.

We present two main results. First, we show that, in stark contrast
to the classical case, independent encoders frequently can take
relatively little advantage of the correlations present between
their states: we prove this via a bound on the achievable rate sum
for sources generating irreducible sets of product vectors. On the
other hand, it is possible to do much better for some special
classes of sources. We show, in particular, that for sources of Bell
states, independent encoders \emph{can} take full advantage of
correlations. The achievable rates, however, are governed by
different formulas than in the classical case, reflecting the
quantum nature of the correlations in the input states.

The paper is structured as follows. Section \ref{sec:defn} gives a
formal definition of the distributed compression problem and shows
how questions about cloning, imprinting~\cite{KoashiI01} and quantum
error correction can be formulated in that framework. It also gives
a statement of the classical theorem governing distributed
compression due to Slepian and Wolf before summarizing previous work
on the quantum version. Section \ref{sec:nogo} contains the
statement and proof of our tighter bound for irreducible sources of
product states. Section \ref{sec:Bell} finds the achievable rate
region for sources generating Bell states. Section
\ref{sec:examples} then provides some further examples, where it
seems likely that the optimal rates lie somewhere between full
utilization of correlations and no utilization at all. We end with a
discussion and some open problems.

We use the following conventions throughout the paper. If
$\cE_{AB} = \{ p_i, \ph_i^{AB} \}$ is an ensemble of bipartite
states then we write $\cE_A$ for the ensemble $\{ p_i, \ph_i^A \}$
of reduced states on system $A$. Sometimes we omit subscripts (or
superscripts) labelling subsystems, in which case the largest
subsystem on which the ensemble (or state) has been defined should
be assumed: $\cE = \cE_{AB}$ and $\ph_i = \ph_i^{AB}$. We identify
states with their density operators and if $\ket{\ph}$ is a pure
state vector, we use the notation $\ph = \proj{\ph}$ for its
density operator. The function $S(\rho)$ is the von Neumann
entropy $S(\rho) = -\Tr \rho \log \rho$ and $S(\cE)$ the von
Neumann entropy of the average state of the ensemble $\cE$.
Functions like $S(A|B)_\r$ and $S(A:B|C)_\r$ are defined in the
same way as their classical counterparts:
\begin{equation}
S(A:B|C)_\r = S(\r^{AC}) + S(\r^{BC}) - S(\r^{ABC}) - S(\r^C),
\end{equation}
for example. $\chi(\cE)$ is
the Holevo $\chi$ quantity of $\cE$~\cite{Holevo73b}.
Throughout, $\log$ and $\exp$ are taken base $2$.

\section{Definition and examples} \label{sec:defn}

We now give a more formal definition of the distributed compression
problem. For convenience, our definition will refer to the case of
two encoders, henceforth known as Alice and Bob. The extension to
any finite number of parties is straightforward. Our receiver will
be named Charlie. Consider an ensemble of bipartite quantum states
$\cE_{AB} = \{ p_i, \ket{\ph_i}^{AB} \}$ on a finite-dimensional
Hilbert space $\cH_{AB} = \cH_A \ox \cH_B$ and the product ensemble
$\cE^{\ox n} = \{ p_{i^n}, \ket{\ph_{i^n}}^{AB} \}$ on
$\cH_{AB}^{\ox n}$, where
\begin{eqnarray*}
i^n &=& i_1 i_2 \dots i_n, \\
p_{i^n} &=& p_{i_1} p_{i_2} \dots p_{i_n} \quad \mbox{and} \\
\ket{\ph_{i^n}} &=&
    \ket{\ph_{i_1}} \ox \ket{\ph_{i_2}} \ox \dots \ox \ket{\ph_{i_n}}.
\end{eqnarray*}
A source provides Alice and Bob with the state $\ket{\ph_{i^n}}$,
drawn with probability $p_{i^n}$. Alice and Bob then perform their
respective encoding operations $E_A$ and $E_B$. These are quantum
operations, that is, completely positive, trace-preserving (CPTP)
maps, with outputs on quantum systems $C_A$ and $C_B$ of
dimensions $d_A$ and $d_B$, respectively. The joint encoding
operation is $E_A \ox E_B$ since Alice and Bob are required to act
independently. The systems $C_A$ and $C_B$ are then sent to
Charlie, who performs a decoding operation $D$, again a CPTP map,
producing the output state $\tilde{\ph}_{i^n} = D \circ (E_A \ox
E_B)(\ph_{i^n})$. We say the \emph{encoding-decoding scheme} has
fidelity $1 - \e$ if
\begin{equation}
\label{eq:fidelity-crit}
\sum_{i^n} p_{i^n} \bra{\ph_{i^n}} \tilde{\ph}_{i^n} \ket{\ph_{i^n}}
\geq 1 - \e
\end{equation}
and that $(R_A,R_B)$ is an \emph{achievable rate pair} if for all
$\d,\e > 0$ there exists an integer $N$ such that for all $n > N$
there is an encoding-decoding scheme with fidelity $1-\e$ satisfying
\begin{equation}
\frac{1}{n} \log d_A \leq R_A + \d
\quad \mbox{and} \quad
\frac{1}{n} \log d_B \leq R_B + \d.
\end{equation}
This scenario is formulated in analogy to the asymptotically
lossless setting of classical block compression, as opposed to
lossless variable-length coding.

We remark here that we may easily allow Alice and Bob the use of
prior shared randomness, without affecting any of our conclusions.
Indeed, randomness is unnecessary, as a look at the fidelity
criterion Eq.~(\ref{eq:fidelity-crit}) shows: the fidelity is an
ensemble expectation of quantities linear in the output state
$\tilde{\varphi}_{i^n}$. Hence the fidelity of a randomized scheme,
regardless of whether it uses shared or private randomness, is the
average of fidelities of the schemes obtained by picking particular
instances of the random data. So, at least one of the
randomness-free schemes has a fidelity at least as good as the
randomized version.

%

The classical correlated source compression problem has a beautiful
solution, due to Slepian and Wolf \cite{SlepianW73}.  This remarkable
theorem shows that Alice and Bob can \emph{always} take advantage of any
correlations that exist between their data.

\begin{figure}
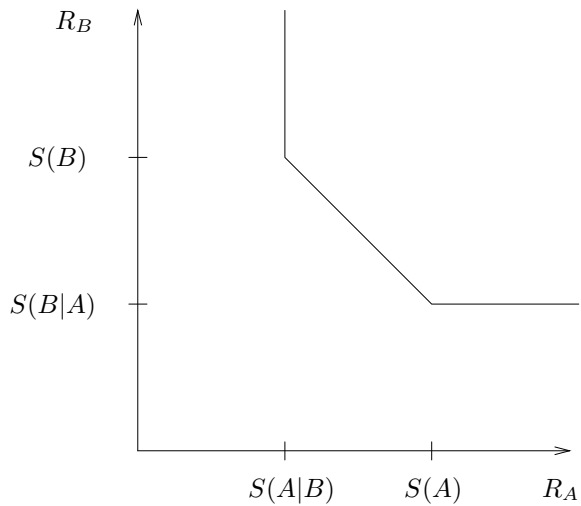

\begin{center}
\input slepwolfregion.eepic
\end{center}
\caption[Slepian-Wolf rate region]{Achievable rate region for
Slepian-Wolf encoding.}
\label{Cfig:slepwolf}
\end{figure}

\begin{theorem}[Slepian-Wolf \cite{SlepianW73}. See also \cite{CoverT}, p.~407]
\label{Cthm:slepianWolf} Let $\cE_{AB} = \{p_i,
\ket{\ph_i}_A\ket{\ps_i}_B \}$ such that $|\braket{\ph_i}{\ph_j}|,
|\braket{\ps_i}{\ps_j}| \in \{0,1\}$. Then $(R_A,R_B)$ is an
achievable rate pair if and only if
\begin{eqnarray}
R_A + R_B &\geq& S(A,B) \label{Ceqn:slepian1}\\
R_A &\geq& S(A|B) \label{Ceqn:slepian2} \\
R_B &\geq& S(B|A) \label{Ceqn:slepian3}.
\end{eqnarray}
\end{theorem}
The entropies here and in our subsequent theorems are taken with
respect to the average state of the ensemble $\cE_{AB}$. We will
refer to inequalities (\ref{Ceqn:slepian1})-(\ref{Ceqn:slepian3})
as the Slepian-Wolf bounds. Note that by time sharing and resource
wasting, achievability of the region defined by the Slepian-Wolf
bounds follows from the achievability of just two rate points:
$(S(A),S(B|A))$ and $(S(A|B),S(B))$. The region is depicted in
Figure \ref{Cfig:slepwolf}.

It is straightforward to show that the Slepian-Wolf bounds hold for
all sources of quantum states~\cite{Winter99,Hayden01} but we will
see in Section \ref{sec:nogo} that in the general case they are
freqently not achievable. In fact, achievability of the Slepian-Wolf
bounds appears to be a singular phenomenon. Nonetheless, Devetak and
Winter have generalized the coding portion of the Slepian-Wolf
theorem to the situation where the states given to one party, say
Alice, are quantum mechanical while those given to the other party
are classical, meaning pure and perfectly distinguishable.  For such
a source, they show that $(S(A),S(B|A))$ is an achievable rate
pair~\cite{DevetakW02}. (In section \ref{subsec:hybrid} we will
combine the technique they used with a type of superdense coding to
develop a coding procedure for partially entangled states.) Whether
the point $(S(A|B),S(B))$ is achievable in their scenario remains
unknown.

\begin{figure}
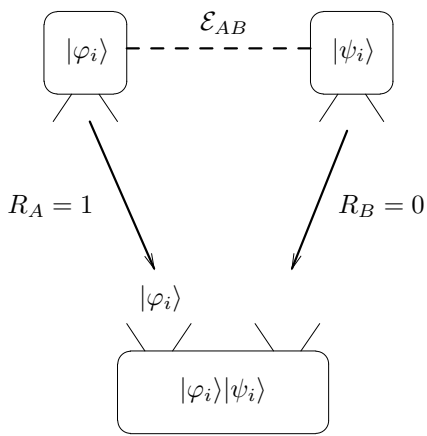

\begin{center}
\input extract.eepic
\end{center}
\caption[Measurement without disturbance as distributed
compression] {Measurement without disturbance as distributed
compression.  This time the encoders are given the states
$\ket{\ph_i}$ and $\ket{\ps_i}$, and the decoder attempts to
produce $\ket{\ph_i}\ket{\ps_i}$.}
\label{Cfig:extract}
\end{figure}

\begin{example}\exampleTitle{Cloning and information-disturbance}
Let's move on to a purely quantum mechanical scenario, in which we will
be able to relate the distributed compression problem to no-cloning and
information-disturbance ideas.  Suppose that
the source generates pairs $\ket{\ph}\ket{\ph} \in \CC^2 \ox \CC^2$
according to the uniform distribution over qubit states.
If Alice is given a noiseless quantum channel with a rate of one qubit
per signal state to Charlie while Bob is given no channel at all, then perfect
reconstruction of the input by Charlie is simply cloning.
This situation is illustrated in Figure \ref{Cfig:clone}.
In the approximate setting, the rate pair $(1,0)$ is achievable if and
only if there exists a sequence of CPTP maps $D_n$ such that
\begin{equation}
\lim_{n\rar\infty} \int
    \bra{\ph_1 \ph_2 \cdots \ph_n} D_n(\ket{\ph_1 \ph_2 \dots \ph_n})
    \ket{\ph_1 \ph_2 \cdots \ph_n} \; d\ph_1 d\ph_2 \cdots d\ph_n = 1.
\end{equation}

\begin{figure}
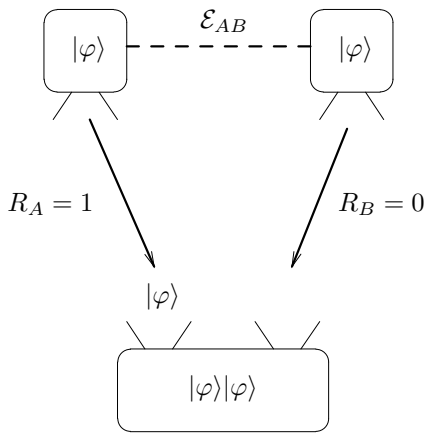

\begin{center}
\input clone.eepic
\end{center}
\caption[Cloning as distributed compression]{Cloning as
distributed compression. Solid lines represent noiseless quantum
channels and dashed lines correlation in the ensemble $\cE_{AB}$.
The encoders are each given a copy of $\ket{\ph}$ while the
decoder tries to produce the state $\ket{\ph}\ket{\ph}$.}
\label{Cfig:clone}
\end{figure}

Similarly, if we replace the uniform ensemble over states
$\ket{\ph}\ket{\ph}$ by some other ensemble
$\{p_i,\ket{\ph_i}_A\ket{\ps_i}_B\}$ and again do not give Bob any
capacity to communicate with Charlie, then studying distributed
compression is simply an information-disturbance problem.  A
graphical depiction is given in Figure \ref{Cfig:extract}. On the
other hand, if Alice is given a full qubit's worth of capacity and
Bob is given some capacity greater than zero but less than a full
qubit, then we are in the regime of information-disturbance
relations with prior correlation~\cite{Hayden01}, since we can
assume that Charlie receives a state of the form $\proj{\ph_{i^n}}
\ox \r_{i^n}$ for some density operator $\r_{i^n}$ and would like to
use a CPTP map to convert it to a state close to
$\ket{\ph_{i^n}}\ket{\ps_{i^n}}$.
\end{example}

\begin{example}\exampleTitle{Erasure codes}
Our final example hints that a full theory of the distributed
compression of entangled states may be related to the analysis of
quantum error correcting codes. Consider the following states:
\begin{eqnarray}\label{eqn:erasure-cw}
\ket{\psi_{00}} &=& \frac{1}{\sqrt{2}}(\ket{0000} + \ket{1111})
\nonumber\\
\ket{\psi_{01}} &=& \frac{1}{\sqrt{2}}(\ket{0011} + \ket{1100})
\nonumber\\
\ket{\psi_{10}} &=&  \frac{1}{\sqrt{2}}(\ket{0101} + \ket{1010})
\nonumber\\
\ket{\psi_{11}} &=&  \frac{1}{\sqrt{2}}(\ket{1001} + \ket{0110}).
\end{eqnarray}
Let $\cE_{AB}$ be the uniform ensemble for the subspace they span,
giving Alice the first two qubits and Bob the last two.
The subspace is, in fact, a type of quantum error correcting code
known as an \emph{erasure code}, capable of correcting for one
error at a known position~\cite{GrasslBP97,VaidmanGW96}. Thus,
$(2,1)$ is an achievable rate pair: Alice sends all of her qubits
while Bob throws away half of his. Meanwhile, the Slepian-Wolf
bounds only require that $R_A + R_B \geq 2$ with no conditions on
$R_A$ and $R_B$ individually. Whether the pair $(2,1)$ is optimal,
then, is actually a question about the approximate performance of
a quantum error correcting code.
\end{example}

\section{A bound for irreducible product state sources}
\label{sec:nogo}

The case of irreducible product state ensembles provides what is
perhaps the most striking example of the unattainability of the
Slepian-Wolf conditions. A set $\mathcal S$ of state vectors is
called \emph{reducible} if its elements fall into two or more
orthogonal subspaces. Otherwise $\mathcal S$ is called
\emph{irreducible}. For more details on the definition and some of
its equivalent formulations, see~\cite{BarnumHJW01}
and~\cite{Hayden01}. Intuitively, an irreducible set of state
vectors is one for which all non-trivial measurements induce at
least some disturbance. We say that an ensemble is irreducible if
the corresponding underlying set of states is. The main result of
this section is a lower bound on the attainable rate sums
$R_A+R_B$ for irreducible ensembles.
We will use two results that have been proved
elsewhere~\cite{BarnumHJW01} which express the fact that an
irreducible ensemble which some quantum operation leaves almost
invariant cannot leak much quantum information to the environment
of the map. These statements can be thought of as approximate and
asymptotic formulations of the no-cloning and
information-disturbance principles.

\begin{lemma}[Barnum et al.~\cite{BarnumHJW01}, Lemma 6.1]
\label{Clemma:irred} Suppose that $\cE = \{ p_i, \ket{\sigma_i}
\}$ is an irreducible ensemble with $K$ states. Suppose that the
states $\ket{\sigma_i}$ are provided in a register A with state
space $\cH_{A}$ and let register B be an ancilla with state space
$\cH_{B}$ (with Hilbert space dimensions $d_A$ and $d_B$); we will
refer to B as the environment. Let
\begin{equation}\begin{split}
  \label{gamma}
  \Gamma : \cH_A\otimes \cH_B
                   &\longrightarrow \cH_A\otimes \cH_B \\
           \ket{\sigma_i}_{A} \ket{0}_B
                   &\longmapsto \ket{\xi_i}_{AB}
\end{split}\end{equation}
be a unitary map such that
$$\sum_i p_i F\bigl( \ket{\sigma_i}, \Tr_B \proj{\xi_i} \bigr) = 1-\epsilon.$$
Let  $\{ p_i, \rho_i = \Tr_A \proj{\xi_i} \}$ be the environment
ensemble and let
\[ \chi\bigl( \{ p_i, \rho_i \} \bigr)
     = S\left( \sum_i p_i \rho_i \right) - \sum_i p_i S(\rho_i ) \]
be the Holevo quantity of the environment. Then if $\cE$ is kept
fixed, but $\epsilon$, $\Gamma$ and $d_B$ are allowed to vary, we
have $\chi \leq f(\epsilon )$ where the function $f$ satisfies
$f(\epsilon ) \rightarrow 0$ as $\epsilon \rightarrow 0$. In fact we
may take $f(\epsilon ) = \alpha\sqrt{\epsilon} +
\beta\sqrt{\epsilon} \log \sqrt{\epsilon}$ where $\alpha$ and
$\beta$ are constants.
\end{lemma}

\begin{proposition}[Barnum et al.~\cite{BarnumHJW01}, Lemma 6.4]
\label{Cprop:info-disturbance}
Consider, for $\epsilon > 0$,
\begin{enumerate}
  \item an ensemble $\cE = \cE_1\otimes \ldots \otimes \cE_{n}$,
    where each $\cE_i$ is an irreducible
    ensemble on a state space of dimension at most $k$ and with at most $K$
    signal states;
  \item an encoding-decoding scheme $(E,D)$ on $\cE$ with average
    fidelity $\geq 1-\epsilon$,
    leaving the environment in a state $\rho_{i^n}$ for input state
    labelled by $i^n = i_1\ldots i_n$.
\end{enumerate}
Then $\frac{1}{n}\chi\bigl( \{p_{i^n}, \rho_{i^n}\} \bigr) <
g(\epsilon)$ where $g$ is a function satisfying $g(\epsilon)
\rightarrow 0$ as $\epsilon \rightarrow 0$. Hence the amount of
information per position tends to zero as the fidelity tends to 1,
for large block lengths $n$.
\end{proposition}

With these tools, we can prove our main result:

\begin{theorem}
\label{Cthm:irred} Let $\cE_{AB} =
\{p_i,\ket{\ph_i}_A\ket{\ps_i}_B \}$ be an irreducible ensemble of
product states. Then a necessary condition for the rate pair
$(R_A,R_B)$ to be achievable for $\cE_{AB}$ is that
\begin{equation} \label{Ceqn:unitaryIrred}
R_A + R_B \geq \frac{ S(\cE_A) + S(\cE_B) + S(\cE_{AB}) }{2}.
\end{equation}
\end{theorem}

\begin{proof}
The basic idea is that if $(R_A,R_B)$ fail to satisfy
Eq.~(\ref{Ceqn:unitaryIrred}), then there is not enough room in
the compressed data to absorb all the distinguishability present
in the input. Some must, therefore, be left behind in the
environments of Alice and Bob. The amount of distinguishability
allowed there, however, is governed by
Proposition~\ref{Cprop:info-disturbance}.

Suppose that, for some $\d,\e > 0$, Alice and Bob have a distributed
encoding-decoding scheme $(E_A \ox E_B,D)$ for blocks of size $n$,
with $\smfrac{1}{n} \log d_A \leq R_A +\d$, $\smfrac{1}{n} \log d_B
\leq R_B + \d$ and fidelity $1-\e$. There exists a unitary extension
of Alice's encoding operation $E_A$ in which the output Hilbert
space factors as $\cH_A = \cH_{W_A} \ox \cH_{C_A}$, where
$\cH_{W_A}$ is waste and $\cH_{C_A}$ represents her noiseless
quantum channel. Thus, $E_A(\rho) = \Tr_{W_A} U_A (\rho \ox
\proj{0}) U_A^\dagger$ for some unitary $U_A$ and fixed ancilla
state $\proj{0}$ on $W_A$. Likewise, we can factor Bob's Hilbert
space as $\cH_B = \cH_{W_B} \ox \cH_{C_B}$ and write $E_B(\rho) =
\Tr_{W_B} U_B(\rho \ox \proj{0}) U_B^\dagger$. Now, let $\rho^A =
\sum_{i^n} p_{i^n} \proj{\ph_{i^n}}$ and $\rho^B = \sum_{i^n}
p_{i^n} \proj{\psi_{i^n}}$. Then, by the subadditivity and unitary
invariance of the von Neumann entropy, we find
\begin{equation}
  \label{eq:entropybound}
  S(\cE_A^{\ox n}) = S(\rho^A)
                   = S\Big( U_A(\rho^A\ox\proj{0})U_A^\dagger \Big)
                \leq S(\rho^{W_A}) + S(\rho^{C_A}),
\end{equation}
where
\begin{equation*}
  S(\rho^{W_A}) = S\big( \sum_{i^n} p_{i^n} \Tr_{C_A}\bigl[ U_A(\proj{\ph_{i^n}}\ox\proj{0})U_A^\dagger
  \bigr] \big)
\end{equation*}
is the average density operator for the reduced state of Alice's
waste area and where
\begin{equation*}
  S(\rho^{C_A}) = S\big( \sum_{i^n} p_{i^n} \Tr_{W_A}\bigl[
U_A(\proj{\ph_{i^n}}\ox\proj{0})U_A^\dagger \bigr] \big).
\end{equation*}
If we define
\begin{align*}
  \rho_{i^n}^{W_A} &= \Tr_{C_A}\bigl[ U_A(\proj{\ph_{i^n}}\ox\proj{0})U_A^\dagger \bigr]
                                                                     \ \text{ and} \\
  \rho_{i^n}^{C_A} &= \Tr_{W_A}\bigl[ U_A(\proj{\ph_{i^n}}\ox\proj{0})U_A^\dagger \bigr],
\end{align*}
and note that $n (R_A +\d) \geq S(\rho^{C_A})$, since $\rho^{C_A}$
is a state on a Hilbert space of dimension at most
$2^{n(R_A+\d)}$, we can then use Eq.~(\ref{eq:entropybound}) to
conclude that
\begin{align}
  R_A + \d &\geq S(\cE_A) - \frac{1}{n}
                       \left( \chi\bigl( \{p_{i^n},\rho_{i^n}^{W_A}\} \bigr)
                                        - \sum_{i^n} p_{i^n} S(\rho_{i^n}^{W_A}) \right) \nonumber\\
      &=    S(\cE_A) - \frac{1}{n}
                       \left( \chi\bigl( \{p_{i^n},\rho_{i^n}^{W_A}\} \bigr)
                                        - \sum_{i^n} p_{i^n} S(\rho_{i^n}^{C_A}) \right),
                                                                       \label{eq:RA-chi}
\end{align}
where in the last line we have used that
$S(\rho_{i^n}^{W_A})=S(\rho_{i^n}^{C_A})$. An analogous inequality
obviously holds for B. At this point, we have come close to
isolating the distinguishability left behind in the Alice waste
area, in the form of $\frac{1}{n}\chi\bigl(
\{p_{i^n},\rho_{i^n}^{W_A}\} \bigr)$, which goes to $0$ as
$n\rar\infty$ and $\epsilon\rar 0$ by
Proposition~\ref{Cprop:info-disturbance}. But our expression also
depends on the average mixedness of the channel states
$\rho_{i^n}^{C_A}$. We can control this through a series of
inequalities that follow from the properties of $\chi$, however:
\begin{equation*}\begin{split}
  \chi\bigl( \{p_{i^n},\rho_{i^n}^{C_A}\} \bigr)
   + \chi\bigl( \{p_{i^n},\rho_{i^n}^{C_B}\} \bigr)
            &\geq \chi\bigl( \{p_{i^n},\rho_{i^n}^{C_A}\ox\rho_{i^n}^{C_B}\} \bigr) \\
            &\geq \chi\bigl( \{p_{i^n},D(\rho_{i^n}^{C_A}\ox\rho_{i^n}^{C_B})\} \bigr) \\
            &\geq S(\cE_{AB}^{\ox n}) - n h(\epsilon),
\end{split}\end{equation*}
where $h(\epsilon)\rar 0$ as $\epsilon\rar 0$. The three
inequalities follow, in order, from the superadditivity of $\chi$
for ensembles of product states~\cite{Holevo79}, the
Lindblad-Uhlmann monotonicity of $\chi$ under quantum channels,
and the Fannes inequality~\cite{Fannes73}. Again using $n (R_A
+\d) \geq S(\rho^{C_A})$ and $n (R_B +\d) \geq S(\rho^{C_B})$,
this inequality implies that
$$\sum_{i^n} p_{i^n}\bigl( S(\rho_{i^n}^{C_A})+S(\rho_{i^n}^{C_B}) \bigr)
                \leq n \big( R_A + R_B - S(\cE_{AB}) + 2\d + h(\epsilon) \big).$$
This, in turn combined with Inequality~(\ref{eq:RA-chi}) and its
counterpart for $R_B$, yields, by invoking
Proposition~\ref{Cprop:info-disturbance},
\begin{equation*}
  2(R_A + R_B) \geq S(\cE_A) + S(\cE_B) + S(\cE_{AB})
                             - 4\d - 2 g(\epsilon) - h(\epsilon),
\end{equation*}
and we are done.
\end{proof}

\section{Optimal compression for sources of Bell states}
\label{sec:Bell}

The result of the previous section, that distributed compression
of irreducible ensembles of product states generically cannot take
full advantage of classical correlations, may be somewhat
discouraging.
Fortunately, this is not quite the end of the story.
In this section we consider mixtures of Bell states. The quantum
correlations present in the ensemble allow us to use a variation
on the hashing protocol for purifying EPR pairs
\cite{BennettDSW96}, combined with a type of superdense coding.
This protocol is fully efficient, in the sense that the total
number of qubits communicated matches the Schumacher bound for
the joint ensemble. We will show the following:
\begin{theorem}\label{Cthm:Bell}
Let
\begin{equation}
\cE_{AB} = \left\{ \begin{array}
  {r@{, \quad}l}
  p_1 & \ket{\phi^+} = \frac{1}{\sqrt{2}}( \ket{00} + \ket{11} ) \\
  p_2 & \ket{\phi^-} = \frac{1}{\sqrt{2}}( \ket{00} - \ket{11} ) \\
  p_3 & \ket{\psi^+} = \frac{1}{\sqrt{2}}( \ket{01} + \ket{10} ) \\
  p_4 & \ket{\psi^-} = \frac{1}{\sqrt{2}}( \ket{01} - \ket{10} )
                   \end{array} \right\}
\end{equation}
be an ensemble of Bell pairs, and let $H = H(p_1, p_2, p_3,
p_4)$. Then the rate pair $(R_A,R_B)$ can be achieved by distributed
compression
if and only if
\begin{equation} \label{Ceqn:Bellrate}
R_A \geq H/2 \quad \mbox{and} \quad R_B \geq H/2.
\end{equation}
\end{theorem}

\subsection{Proof of achievability}

While the states in the ensemble are highly entangled,
they are also mutually orthogonal. So, while
the ensemble $\cE_{AB}$
is highly quantum mechanical from the points of view of Alice and Bob,
it is classical from the point of view of the decoder, whose operations are
not encumbered by any locality constraints. Our protocol makes use of this
obseration in an essential way: Alice and Bob will perform a series of
local unitary operations before sending some fraction of their Bell
pairs to Charlie, who will then perform a measurement to establish
the identity of the states he has received. By appropriate choices
of the local operations, all the information about the input can
be hashed into the identity of the state sent to Charlie.

A Bell pair can be labelled by a pair of bits. We will follow the
convention of Ref.~\cite{BennettDSW96}, in which the
Bell pair state $\ket{0}\ket{y_1} + (-1)^{y_2} \ket{1}\ket{1-y_1}$ is
represented by the label $(y_1,y_2)$. This labelling has the property
that given two Bell pairs
described by $(y_1,y_2)$ and $(z_1,z_2)$, local unitary operations suffice
to add $z_1$ or $z_2$ to either of $y_1$ or $y_2$. For example,
a bilateral CNOT can be used to implement the transformation
\begin{equation}
\label{eqn:bitmap}
(z_1, z_2), (y_1, y_2) \mapsto (z_1 + y_1, z_2), (y_1, y_2+z_2).
\end{equation}
(Note, however, that although the operation succeeds in adding
$z_2$ to $y_2$, there is an unavoidable ``backaction'' on $y_1$.)
With this convention, a sequence of $n$ Bell pairs can be
described by a $2n$-bit string, which we shall denote by
$\vec{x}^n$. This string, in turn, can be considered as a
concatenation of two strings $\vec{x}^C$ and $\vec{x}^W$ that are
$2 m$ and $2 (n - m)$ bits long, respectively. $\vec{x}^C$ will
represent the bits that Alice and Bob send through the channel to
the decoder, and $\vec{x}^W$ will represent the bits that are
thrown away.

We will use a protocol in which Alice and Bob share $2 m$ random
$2 (n-m)$-bit-long strings $\vec{s}(k)$, where $k$ ranges from $1$ to $2(n-m)$;
the necessity of sharing randomness can be removed
from the final protocol by observing that the average fidelity of
the protocol is the probability expectation (over the shared
randomness) of the average fidelities of schemes with the
value of the shared radomness fixed.

The protocol is much
like hashing and consists of $2m$ rounds of the following
procedure. In the $k$th round, given the random strings above, Alice
and Bob replace $x^C_k$ with ${x^C_k}' = x^C_k + \vec{s}(k) \cdot
\vec{x}^W$ using local operations as discussed above. The effect of
these operations will be to perform $2 m$ random ``bit masks''
on the string
that is ultimately measured by Charlie, who therefore extracts the
parity of a random subset of bits.  After every two rounds $2j$ and
$2j+1$ (where $j$ ranges from $1$ to $m$), Alice and Bob put the Bell
pair described by the bits $x^{C'}_{2j}$ and $x^{C'}_{2j+1}$
aside. Finally, they send all $m$ pairs to Charlie, who measures it in
the Bell basis to ascertain $x^{C'}_{2j}$ and $x^{C'}_{2j+1}$.

We wish to determine the minimal $m$ such that Charlie
can decode the original
$n$ pairs with near-vanishing error probability. Consider two strings
$\vec{x^n}$ and $\vec{y^n}$, where $\vec{x^n}$ is the true initial string.
We will evaluate the probability that $x^n$ and $y^n$ are different but
nonetheless
result in the same $2 m$ decoder outcomes, i.e.,
the decoder cannot uniquely decode the state. Denote the event in which
all the decoder measurements agree for $x^n$ and $y^n$ by $E$. Then
\begin{eqnarray}
\Pr(\vec{x^n} \neq \vec{y^n}, E)
&=& \Pr(\vec{x^n} \neq \vec{y^n}) \Pr(E | \vec{x}^n \neq \vec{y}^n)\nonumber\\
&=& \Pr(\vec{x}^n \neq \vec{y}^n)
    [\Pr(\vec{x}^W \neq \vec{y}^W)
    \Pr (E | \vec{x}^W \neq \vec{y}^W, \vec{x}^n \neq \vec{y}^n)\nonumber\\
   &&+ \Pr(\vec{x}^W = \vec{y}^W) \Pr (E | \vec{x}^W = \vec{y}^W, \vec{x^n} \neq \vec{y^n})]
    \nonumber\\
&=&  \Pr(\vec{x}^n \neq \vec{y}^n)
    [\Pr(\vec{x}^W \neq \vec{y}^W) 2^{-2m} \nonumber\\
     && +  \Pr(\vec{x}^W = \vec{y}^W) \Pr(E | \vec{x}^C \neq \vec{y}^C,
\vec{x}^W = \vec{y}^W)],
\end{eqnarray}
where the last equality follows from multiplying by a factor of $1/2$
for every subsequent random bit mask $\vec{s}(k) \cdot \vec{x}^W$ done
by Alice and Bob.

Now, we argue that the second term in the last equality is
zero. Consider the first number $j$ such that the bit $x^C_j$ is not
equal to $y^C_j$. The information that actually gets sent to the decoder
is in fact more complicated than $x^C_j$ because of the random bit
masks. In each case, the bit that gets sent is
\begin{eqnarray}
{x^C_j}' &=& x^C_j + f(x^W_1, x^W_2, ... , x^W_m) + g(x^C_1,...,x^C_{j-1})\nonumber\\
{y^C_j}' &=& y^C_j + f(y^W_1, y^W_2, ... , y^W_m) + g(y^C_1,...,y^C_{j-1}),
\end{eqnarray}
where $f$ takes into account the bit masks, and $g$ takes into account
the backaction due to previous bit masks. But since $\vec{x}^W =
\vec{y}^W$ for that term and $x^C_k = y^C_k$ for $k < j$ by
hypothesis, the $f$ and $g$ functions are equal, and thus ${x^C_j}'
\neq {y^C_j}'$. Then $ \Pr(E | \vec{x}^C \neq \vec{y}^C, x^W = y^W) = 0$, as we
wished to show. This yields
\begin{equation}
\Pr(\vec{x^n} \neq \vec{y^n}, E) \leq 2 ^ {- 2 m}.
\end{equation}
Additionally, we know that a typical set of candidates for the
initial sequence of size $2^{n(H+\delta)}$ members will with probability
greater than $1 - O(\exp(-\delta^2n))$ contain the true initial
sequence $\vec{x}$~\cite{CoverT}. The decoding will then fail only
for two reasons: the true initial sequence is outside the typical
set or it was impossible to uniquely decode based on the measurement
outcome. Therefore,
\begin{equation}
\Pr(\mathrm{failure}) \leq 2^{n(H + \delta) - 2 m} + O(\exp(-\delta^2n)).
\end{equation}
We can see that if $2 m = n(H + 2\delta)$, the error probability
approaches zero. The number of Bell pairs $m$ that must be sent is just
$n$ times the rate at which Alice and Bob must send their qubits:
\begin{equation}
\label{Ceqn:bellrate}
R_A = R_B = H/2.
\end{equation}

\subsection{Proof of optimality}

The rate pair $(H/2,H/2)$ is also optimal: neither rate can be reduced below
$H/2$. The total
number of qubits sent from Alice and Bob to Charlie must be at least
$H$ by the optimality of Schumacher compression. On the other hand,
Alice and Bob's local density operators are independent of the input.
Intuitively,
all information about the identity of the state exists in the correlations
between their systems. As a result, it is impossible to do better than
splitting the total rate equally between them. For comparison's sake,
observe that the Slepian-Wolf bounds in this case are
\begin{equation}
R_A, R_B \geq H - 1 \quad \mbox{and} \quad R_A + R_B \geq H.
\end{equation}
These inequalities do not ensure $R_A, R_B \geq H/2$, so we see that
even here, where it is possible to fully exploit the correlations,
the Slepian-Wolf bounds are insufficient to describe the achievable
rate region. On the other hand, while it is not applicable in this
case, Theorem~\ref{Cthm:irred} would have given the stronger bound
$R_A+R_B \geq \frac{1}{2}(2+H) > H$, which is in fact violated by
our coding theorem.

In order to prove optimality of the given rate pair, it is sufficient
to show that $R_A \geq H/2$ regardless of the size of $R_B$. In what
follows, we can therefore assume that Bob noiselessly transmits all
of his source qubits to Charlie.
We can augment any high-fidelity compression scheme by a
state preparation scheme. Imagine a state preparer,
Peter, who prepares Bell states according to the given distribution before
giving one qubit of each pair to Alice and the other to Bob. Alice
and Bob compress these Bell states as before. If the
average fidelity of the compression scheme is $1 - \epsilon$, we can
think of this augmented state-preparation/compression scheme as
classical communication from Peter to Charlie with average error
probability $\epsilon$. The Fannes inequality~\cite{Fannes73} ensures
that there exists a function $f(\epsilon)$ that approaches zero as
$\epsilon$ approaches zero such that the classical communication rate
from Peter to Charlie, measured in bits, is $H - f(\epsilon)$.

Let us define Peter's state preparation more precisely: for each Bell
state, he can prepare a singlet, give one of the qubits to Bob, and
then act on the other qubit with an appropriate Pauli rotation before
handing it to Alice. Since Bob will give all his qubits to Charlie
perfectly anyway, we can eliminate Bob from consideration and consider
an equivalent picture in which Peter initially shares singlets with
Charlie and encodes his classical information by acting with Paulis
according to a distribution of entropy $H$. In this communication
channel from Peter to Charlie, Alice is the bottleneck: she sends
qubits at rate $R_A$. This rate assisted by entanglement can
simply be thought of as superdense coding; it can result in a
classical transmission rate from Peter to Charlie of at most $2
R_A$. Combining this rate with our other expression for this classical
transmission rate gives
\begin{equation}
2 n R_A \geq n (H - f(\epsilon)).
\end{equation}
Letting $\epsilon \rightarrow 0$ proves that $R_A \geq
H/2$. Switching the roles of Alice and Bob completes the proof.

\section{Further examples} \label{sec:examples}

In this section we present a pair of examples that are designed to
illustrate the range of compression strategies available to
encoders. In each case, as with the optimal Bell pair strategy,
the key is to make make use of orthogonality in the ensemble even
though it is not directly accessible to the encoders.

\subsection{Hidden orthogonality} \label{subsec:hiddenOrthog}

Based on the results of Section \ref{sec:nogo}, one might imagine
that since Alice and Bob must act locally, a system in which both
Alice and Bob's ensembles are \emph{locally} irreducible (and
consisting of pure states) would suffice for Alice and Bob not to
be able to take full advantage of correlations. However, this is
not the case, as we will show in an example that demonstrates that
compressing correlated reducible product sources can involve quite
subtle strategies. This example demonstrates the necessity of
\emph{global} irreducibility in Theorem \ref{Cthm:irred}.

Let $\cE_{AB} = \{1/3, \ket{\ph_i}_A \otimes \ket{\psi_i}_B \}$
where
\begin{eqnarray}
\ket{\ph_1} &=& \ket{0} \\
\ket{\ph_2} &=& \sqrt{\alpha}\ket{0} + \sqrt{1-\alpha}\ket{1} \\
\ket{\ph_3} &=& \ket{1} \\
\ket{\psi_1} &=& \sqrt{1-\beta}\ket{0} + \sqrt{\beta}\ket{1} \\
\ket{\psi_2} &=& \ket{1} \\
\ket{\psi_3} &=& \sqrt{1-\beta}\ket{2} + \sqrt{\beta}\ket{0}
\end{eqnarray}
and both $\a$ and $\b$ are assumed to be small but non-zero. This
ensemble is irreducible from the points of view of $A$ and $B$
individually but is reducible for $AB$. That is, $\cE_A$ and $\cE_B$
are irreducible but $\cE_{AB}$ is not, since
$\ket{\ph_1}\ox\ket{\psi_1} \perp \ket{\ph_3}\ox\ket{\psi_3}$ and
$\ket{\ph_2}\ox\ket{\psi_2} \perp \ket{\ph_3}\ox\ket{\psi_3}$.

\begin{figure}
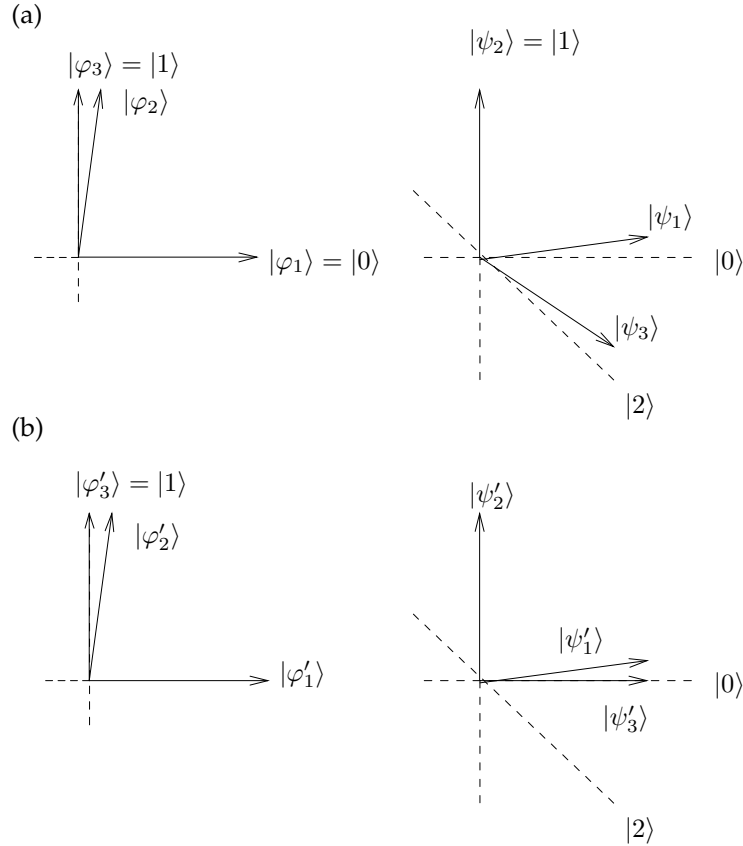

\begin{center}
\input surprise.eepic
\end{center}
\caption[Hidden orthogonality]{Hidden orthogonality: (a) depicts
the ensemble states $\{\ket{\ph_i}\ket{\ps_i}\}$
while (b) shows
the ensemble after Bob has performed the first half of his
compression operation.}
\label{Cfig:surprise}
\end{figure}

The encoder at $A$ simply performs Schumacher compression at the rate
$S(\cE_A) \approx H(2/3)$. The encoder at $B$ begins by projecting
onto $\ket{2}$ and the subspace of states orthogonal to $\ket{2}$,
which we write as $\ket{2}^\perp$. If the outcome is $\ket{2}$, he
sets the state to $\ket{0}$.  This operation has the effect
$\ket{\psi_i} \mapsto \ket{\psi_i'}$ where $\ket{\psi_1'} =
\ket{\psi_1}$, $\ket{\psi_2'}=\ket{\psi_2}$ and $\ket{\psi_3'} =
\ket{0}$.  The effect of the operation is shown in Figure
\ref{Cfig:surprise}.  The encoder then performs Schumacher
compression on the ensemble $\{1/3,\ket{\psi_i'}\}$ at rate
$H(2/3)+f(\beta)$ where $f(\beta)\rar 0$ as $\beta \rar 0$.

The decoder first Schumacher-decompresses the outputs of Alice and
Bob's channels individually. Next he projects onto $\ket{10}$ and
$\ket{10}^\perp$. Notice that
\begin{equation}
\Span(\ket{\ph_1}\ket{\psi_1},\ket{\ph_2}\ket{\psi_2})
                                        \subset \ket{10}^\perp
\end{equation}
and that $\ket{\ph_3}\ket{\psi_3'}=\ket{10}$. Therefore, if the
outcome is $\ket{10}$, he sets the state to
$\ket{\ph_3}\ket{\psi_3}$. Otherwise, he does nothing.

In this way, $\cE_{AB}$ can be compressed to approximately
$(H(2/3),H(2/3))$ qubits per signal.  Emphasizing the need for global,
not just local, irreducibility in Theorem \ref{Cthm:irred}, we can
calculate that for this scheme, $R_A + R_B \approx 2H(2/3) \approx
1.8366$.  On the other hand, the lower bound from the theorem is
\begin{equation}
  \frac{1}{2}\bigl( S(\cE_A)+S(\cE_B)+S(\cE_{AB}) \bigr)
                           \approx \frac{1}{2}H(2/3) + \log 3 \approx 2.0441,
\end{equation}
a rate which is clearly bettered by this example.

Summarizing, Bob performs a locally dissipative operation that can
only be reversed by combining his output with the output of Alice's
channel.  This regime, in which the ensembles are locally
irreducible but globally reducible, seems to provide the greatest
variety of effects and would consequently seem to be the hardest to
solve in general. Indeed, the Bell state example of the previous
section also falls into this category. These types of semi-classical
strategies promise to frequently beat the bounds that apply to fully
irreducible ensembles, but the optimal rates in the general case are
completely unknown.

\subsection{A hybrid strategy} \label{subsec:hybrid}

In this example, we return to the realm of orthogonal entangled
states but without requiring that the states be maximally
entangled.  The compression strategy will combine ideas from the
hidden orthogonality example of section \ref{subsec:hiddenOrthog},
specifically the locally irreversible measurement, and the
protocol for compressing Bell states in Section \ref{sec:Bell}, in
which local unitary transformations were used to ``piggyback''
extra information onto the fraction of states sent to the decoder.

Let $\cE_{AB}$ be an ensemble consisting of two orthogonal
states, $\ket{\ph_0}$ and $\ket{\ph_1}$, in $\CC^2 \ox \CC^2$ occurring with
probabilities $p_0$ and $p_1$, respectively.
By a result of Walgate
et al.~\cite{WalgateSHV00} we may assume without loss of generality that
\begin{eqnarray}
\ket{\ph_0} &=& \a_0 \ket{00} + \b_0 \ket{11} \quad \mbox{and} \\
\ket{\ph_1} &=& \a_1 \ket{01} + \b_1 \ket{10},
\end{eqnarray}
since any other ensemble will be locally equivalent to one of this
type.

As we said, the idea behind this example is to combine two
different strategies. Suppose, given a state $\ket{\ph_i}$ drawn
from $\cE_{AB}$, that Alice performs a projective measurement in
the standard $\{ \ket{0},\ket{1} \}$ basis, whose outcome is
$\ket{j}$. First, observe that if she sends the outcome on to
Charlie and Bob also sends his state to Charlie, then Charlie can
uniquely identify $i$, the identity of the input state. ($i$ is a
function of the parity of the outcomes of local measurements in the
standard basis.) Whenever Alice's measurement outcome is not
independent of the (classical) post-measurement state on Bob's
system, compression of Alice's communication below the rate $H(j)$
will be possible, according to the Slepian-Wolf theorem. Up to
this point, the strategy is effectively classical. To go beyond
Slepian-Wolf, given a state $\ket{\ph_{i^n}}$ drawn from
$\cE_{AB}^{\ox n}$, we will have Alice measure only $n-m$ states,
encoding information about the outcome on the remaining $m$, which
will be sent to Charlie.

Let us estimate the rate achievable using this procedure. Denote
by $q_j$ the probability that Alice gets outcome $j$, by $\o_j^B$
Bob's state given that Alice has measured $j$ and by $\cE'$ the
ensemble $\{ q_j, \o_j^B \}$. Then
\begin{equation}
\o_0^B = \frac{1}{q_0} \sum_i p_i |\a_i|^2 \proj{i} \quad
\mbox{and} \quad
 \o_1^B = \frac{1}{q_1} \sum_i p_i |\b_i|^2
\proj{\neg i}.
\end{equation}
Alice will perform the measurement on the product register $W =
A_1 \dots A_{|W|}$, where $|W| = n - m$. The number of typical
$j^W$ strings will be roughly $\exp(|W|H(\cE'))$. Moreover, that
set will partition into subsets of size roughly
$\exp(|W|\chi(\cE'))$ (and a low-probability remainder) for which
Bob's density operators can be distinguished with negligible
probability of error. Hence, Alice will only need to send
$|W|(H(\cE') - \chi(\cE'))$ bits. This she will do by applying
unitary encodings on her unmeasured states. Denote by $\cE''$ the
ensemble of states $(U \ox I)\r^{AB}(U^\dg \ox I)$, for a set of
unitaries satisfying $\EE \, U \ps U^\dg = I/\dim(A)$ for all states
$\ps$ and $\r^{AB} = \sum_i p_i \ph_i^{AB}$. (For qubits, applying
a random Pauli operator will do.) By the
Holevo-Schumacher-Westmoreland
(HSW) theorem \cite{Holevo98b,SchumacherW97}
this encoding of classical information in quantum states can achieve
the communication rate
\begin{eqnarray}
\chi(\cE'')
&=& S(\cE'') - \EE S((U \ox I)\r^{AB}(U^\dg \ox I)) \\
&=& \log \dim(A)+ S(\r^B) - S(\r^{AB}).
\end{eqnarray}
Thus, requiring that
\begin{equation}
m \chi(\cE'') = (n-m)\left(H(\cE') - \chi(\cE')\right)
\end{equation}
yields a rate $m/n$ for Alice of
\begin{equation}
R_A = \frac{H(\cE')-\chi(\cE')}{H(\cE')-\chi(\cE')+\chi(\cE'')}.
\end{equation}

As strange as this formula looks, it is important to observe that
if $p_i = 1/2$ and $\a_i = \b_i = 1/\sqrt{2}$, we recover the
optimal rate $R_A = 1/2$ from our study of the compression of Bell
states. In our proposal, however, Bob must always send at a rate $R_B =
S(\cE_B)$, which is not optimal in this case.

We will now show more carefully that this procedure actually
works. The argument will essentially just require patching
together known results. The versions we present here are all from
Ref.~\cite{Winter99}. First, we will need the HSW theorem:
\begin{theorem}[Holevo-Schumacher-Westmoreland~\cite{Holevo98b,SchumacherW97}]
\label{Clemma:HSW}
Consider the ensemble $\cE = \{ p_i, \rho_i \}$.  For $0 < \tau < 1$,
$\lambda < 1$, and sufficiently large $n$, there is some $\delta <
K'/\sqrt{n}$ such that the following holds: For a subset $A$ of the
ensemble $\cE ^ {\otimes n}$ such that the total probability of the
states in $A$ is greater than or equal to $\tau$, and a classical
alphabet $M = \{ 1,\ldots, 2^{n\mu} \}$, there exists a code (composed of
a function $f$ that maps elements of $M$ to codestates
$\gamma_k := \rho_{i_1} \otimes ... \otimes \rho_{i_n} \in A$
and an observable $E$ on the
Hilbert space of the codewords) such that the maximum error probability
(defined as ${\rm max}_k \{ 1 - {\rm tr}(\gamma_k E_k): k \in M \}$)
is $\lambda$, and $\mu \geq \chi(\cE) - \delta$.
\end{theorem}
This, in turn, implies the code partition theorem, which we will also use:
\begin{theorem} \label{Clemma:code_partition}
Again, consider the ensemble $\cE = \{ p_i, \rho_i \}$ and the
$n$-block version, $\cE ^ {\otimes n}$.
For any $\lambda, \delta, \eta > 0$ and for
sufficiently large $n$, there exist $m \leq 2 ^ {n(H(\cE) -
\chi(\cE) + 3\delta)}$ many $n$-block codes (as in HSW) with
maximum error probability $\lambda$ and pairwise disjoint ``large''
codebooks $C_i$: $|C_i| \geq 2 ^ {n(\chi(\cE) - 2\delta)}$ such that
$\Pr\left\{\text{state from }\cE ^ {\otimes n}
  \text{ not in }\bigcup_{i=1}^{m} C_i\right\} < \eta$.
\end{theorem}
Finally, the gentle measurement lemma will also be useful. This
result ensures that if Charlie can ascertain Alice and Bob's
states with near-zero chance of error, then he can do so without
causing any significant disturbance. (In this lemma, $\| \cdot
\|_1$ denotes the trace norm.)
\begin{lemma}\label{Clemma:tender_measurement}
Let $\{\rho_a\}, a \in A$
be a family of states, and $E$ an observable indexed by $b \in B$. Let
$\phi: A \rightarrow B$ be a map and let there be $\lambda > 0$ such
that for every $a \in A$, $1 - {\rm tr}(\rho_a E_{\phi(a)}) \leq
\lambda$, i.e., the observable identifies $\phi(a)$ from $\rho_a$ with
maximal error probability $\lambda$. Then the measurement disturbs the
states $\rho_a$ very little: for every $a \in A$, $\| \rho_a - \sum_{b\in
B}\sqrt{E_b} \rho_a \sqrt{E_b} \|_1 \leq \sqrt{8 \lambda} + \lambda$.
\end{lemma}

According to the code partition theorem, for any $\l, \d, \eta > 0$ and
sufficiently large $|W|$, the ensemble $\cE'^{\ox |W|}$ ``partitions''
into at most $\exp(|W|(H(\cE')-\chi(\cE')+3\d))$ codes, each with probability
of error at most $\l$ and containing at least $\exp(|W|(\chi(\cE')-2\d))$
codewords, such that the probability of any state in $\cE'^{\ox n}$ not
lying in any of the codes is less than $\eta$. By the HSW theorem,
for any $\l', \d' > 0$, Alice can find a second code based on $\cE''$
with maximum error
probability $\l'$ containing at least $\exp(m(\chi(\cE'')-2\d''))$
codewords. Therefore, she will be able to send the identity of the
code from the code partition theorem this way provided
\begin{equation}
m ( \chi(\cE'')- 2\d') \geq (n-m)\left( H(\cE') - \chi(\cE') + 3\d \right),
\end{equation}
which gives the same rate we found earlier in our rough estimate.
It remains to show that Charlie can still recover the original
state once he has decoded the piggy-backed information about the
code identity. The probability of error in identifying the code is
bounded above by $\l'$. Let $D$ be the complement of $W$ (the
system sent from Alice to Charlie) and recall that the identity of
the code, call it $k$, is encoded by applying a unitary operator
$U_k \ox I_B$ to the state $\r^{D}$. In reality, however, $\r^{D}$
is an average over input states: $\r^{D} = \sum_{i^{D}} p_{i^{D}}
\ph_{i^{D}}$. Let $\ket{\ph_{i^{D},k}} = (U_k \ox I_B)
\ket{\ph_{i^{D}}}$ and let $\tau_{i^{D},k} = \sum_{k'}
\sqrt{E_{k'}} \ph_{i^{D},j} \sqrt{E_{k'}}$ be Charlie's
post-measurement state. By the gentle measurement lemma,
\begin{eqnarray}
\; \sum_{i^{D}} p_{i^{D}} \| \ph_{i^{D},k} - \tau_{i^{D},k} \|_1
&=&
\Big\| \sum_{i^{D}} p_{i^{D}} \proj{i^{D}} \ox \ph_{i^{D},k}
 - \sum_{i^{D}} p_{i^{D}} \proj{i^{D}} \ox \tau_{i^{D},k} \Big\|_1 \\
&\leq& \sqrt{8 \l'} + \l'.
\end{eqnarray}
Now, the total probability of error on the first $n-m$ states
is bounded above by $\l' + \l + \eta$. On the rest, the decoding
consists of applying $U_{k'}^\dg \ox I_B$, where $k'$ is the measured
code. Noting that $1-F(\r,\s) \leq \smfrac{1}{2}\|\r-\s\|_1$
for any states $\r$ and $\s$~\cite{Fuchsv99},
we find that the average fidelity goes to one as $\l'$ goes to zero.
Thus, the overall average fidelity goes to one as $\l' + \l + \eta$
goes to zero.

\section{Discussion} \label{sec:discussion}

We have studied the problem of performing distributed compression
on a source of correlated quantum states. For some sources, namely
sources of an irreducible set of product states, we find that it
is much harder to exploit correlations in a compression protocol
than would be suggested by the classical Slepian-Wolf theorem. We
did not attempt to find a coding strategy matching the bound of
our Theorem~\ref{Cthm:irred}. Indeed, since its first formulation
in~\cite{Hayden01}, we found the lower bound so odd that none of
us even suspected that it might be tight. (It did lead us to
develop some unwarranted pessimism about the problem, however. We
included in an earlier preprint version of this paper the
erroneous assertion that local Schumacher compression is optimal
for compression of irreducible product state sources. That is true
if only unitary decoding operations are permitted but not in
general.) In any case, the very recent solution of the quantum
Slepian-Wolf problem with free classical
side-communication~\cite{HorodeckiOW04}, which occurred roughly a
year after initial posting of the present paper, and coding
results obtained thereafter for our model without classical
communication~\cite{ADHW05}, show that the rate pair $\bigl(
S(\cE_A), \frac{1}{2}\bigl( S(\cE_B)+S(\cE_{AB})-S(\cE_A) \bigr)
\bigr)$ is indeed universally achievable. In other words, quite
surprisingly, our bound of Theorem~\ref{Cthm:irred} is tight in
the sense that it gives the complete rate region for irreducible
ensembles.

For sources of Bell states, on the other hand, we demonstrated an
optimal method of compression based on the hashing protocol for
entanglement distillation, that fully exploits the quantum
correlations between the two encoders. Nonetheless, the optimal
rate region is not captured by the direct quantum analog of the
classical result due to Slepian and Wolf, nor by our tighter
bound. We also provided some other examples to illustrate the
types of protocols that might occur in an intermediate regime,
where it appears possible to exploit some of the correlations
between the local sources but not all.

Thus, as compared to the classical version of the problem, we find
a bewildering array of different strategies and achievable rates
that are not easily synthesized into a single formula. Finding such
a formula and a uniform approach to the problem integrating
all possible ensembles remains an important open problem.

\medskip
{\bf A Postscript:} We note that the more recent
studies~\cite{HorodeckiOW04,ADHW05} change the model slightly: the
source is described not by an ensemble but by a density operator.
Compression has to succeed for all possible decompositions of that
density operator into pure state ensembles, which is equivalently
described by saying that the purification of the source density
operator has to be preserved with high (entanglement) fidelity. It
turns out that in this model one can show, regardless of the
source, that $R_B \geq \frac{1}{2}\bigl(
S(\cE_B)+S(\cE_{AB})-S(\cE_A) \bigr)$, and analogously for
$R_A$~\cite{Oppenheim05}. Hence, even in this related but
different model we are rather close to understanding the full rate
region.

\subsection*{Acknowledgments}
The authors would like to thank Dominic Berry, Sumit Daftuar, Igor
Devetak, Michal Horodecki, Debbie Leung, Jonathan Oppenheim, Hideo
Mabuchi, John Smolin and Jon Thomas Yard for enjoyable and helpful
conversations. The authors acknowledge the support of the US
National Science Foundation through grant no. EIA-0086038. CA is
also supported by an Institute for Quantum Information fellowship,
PH by the Sherman Fairchild Foundation, CIAR, NSERC and the Canada
Research Chairs program, ACD by the Caltech MURI Center for Quantum
Networks (DAAD19-00-1-0374) and AW by the U.K. Engineering and
Physical Sciences Research Council.

\bibliographystyle{unsrt}
\bibliography{distrib}

\end{document}

%% file: slepwolfregion.eepic
\setlength{\unitlength}{0.00063333in}
\begingroup\makeatletter\ifx\SetFigFont\undefined
\def\x#1#2#3#4#5#6#7\relax{\def\x{#1#2#3#4#5#6}}%
\expandafter\x\fmtname xxxxxx\relax \def\y{splain}%
\ifx\x\y   
\gdef\SetFigFont#1#2#3{%
  \ifnum #1<17\tiny\else \ifnum #1<20\small\else
  \ifnum #1<24\normalsize\else \ifnum #1<29\large\else
  \ifnum #1<34\Large\else \ifnum #1<41\LARGE\else
     \huge\fi\fi\fi\fi\fi\fi
  \csname #3\endcsname}%
\else
\gdef\SetFigFont#1#2#3{\begingroup
  \count@#1\relax \ifnum 25<\count@\count@25\fi
  \def\x{\endgroup\@setsize\SetFigFont{#2pt}}%
  \expandafter\x
    \csname \romannumeral\the\count@ pt\expandafter\endcsname
    \csname @\romannumeral\the\count@ pt\endcsname
  \csname #3\endcsname}%
\fi
\fi\endgroup
{\renewcommand{\dashlinestretch}{30}
\begin{picture}(4888,4051)(0,-10)
\path(975,1614)(1125,1614)
\path(975,2814)(1125,2814)
\path(2250,489)(2250,339)
\path(3450,489)(3450,339)
\path(2250,4014)(2250,2814)(3450,1614)(4650,1614)
\path(1050,414)(4575,414)
\path(4455.000,384.000)(4575.000,414.000)(4455.000,444.000)
\path(1050,414)(1050,4014)
\path(1080.000,3894.000)(1050.000,4014.000)(1020.000,3894.000)
\put(4350,39){$R_A$}
\put(3225,39){$S(A)$}
\put(1950,39){$S(A|B)$}
\put(150,2739){$S(B)$}
\put(0,1539){$S(B|A)$}
\put(375,3864){$R_B$}
\end{picture}
}

%% file: clone.eepic
\setlength{\unitlength}{0.00063333in}
\begingroup\makeatletter\ifx\SetFigFont\undefined%
\gdef\SetFigFont#1#2#3#4#5{%
  \reset@font\fontsize{#1}{#2pt}%
  \fontfamily{#3}\fontseries{#4}\fontshape{#5}%
  \selectfont}%
\fi\endgroup%
{\renewcommand{\dashlinestretch}{30}
\begin{picture}(2874,3837)(0,-10)
\put(117,2892){\arc{210}{1.5708}{3.1416}}
\put(117,3357){\arc{210}{3.1416}{4.7124}}
\put(582,3357){\arc{210}{4.7124}{6.2832}}
\put(582,2892){\arc{210}{0}{1.5708}}
\path(12,2892)(12,3357)
\path(117,3462)(582,3462)
\path(687,3357)(687,2892)
\path(582,2787)(117,2787)
\path(237,2787)(87,2562)
\path(462,2787)(612,2562)
\put(237,3087){$\ket{\ph}$}
\put(2292,2892){\arc{210}{1.5708}{3.1416}}
\put(2292,3357){\arc{210}{3.1416}{4.7124}}
\put(2757,3357){\arc{210}{4.7124}{6.2832}}
\put(2757,2892){\arc{210}{0}{1.5708}}
\path(2187,2892)(2187,3357)
\path(2292,3462)(2757,3462)
\path(2862,3357)(2862,2892)
\path(2757,2787)(2292,2787)
\path(2412,2787)(2262,2562)
\path(2637,2787)(2787,2562)
\put(2412,3087){$\ket{\ph}$}
\put(717,117){\arc{210}{1.5708}{3.1416}}
\put(717,582){\arc{210}{3.1416}{4.7124}}
\put(2232,582){\arc{210}{4.7124}{6.2832}}
\put(2232,117){\arc{210}{0}{1.5708}}
\path(612,117)(612,582)
\path(717,687)(2232,687)
\path(2337,582)(2337,117)
\path(2232,12)(717,12)
\path(687,912)(837,687)
\path(1212,912)(1062,687)
\path(1737,912)(1887,687)
\path(2262,912)(2112,687)
\thicklines
\path(387,2562)(912,1362)
\thinlines
\path(836.417,1459.914)(912.000,1362.000)(891.386,1483.963)
\thicklines
\path(2487,2487)(2037,1362)
\thinlines
\path(2053.713,1484.559)(2037.000,1362.000)(2109.421,1462.275)
\thicklines
\dashline{120.000}(687,3162)(2187,3162)
\put(837,1062){$\ket{\ph}$}
\put(1287,3312){$\cE_{AB}$}
\put(2412,1812){$R_B = 0$}
\put(-287,1812){$R_A = 1$}
\put(1187,312){$\ket{\ph}\ket{\ph}$}
\end{picture}
}

%% file: extract.eepic
\setlength{\unitlength}{0.00063333in}
\begingroup\makeatletter\ifx\SetFigFont\undefined%
\gdef\SetFigFont#1#2#3#4#5{%
  \reset@font\fontsize{#1}{#2pt}%
  \fontfamily{#3}\fontseries{#4}\fontshape{#5}%
  \selectfont}%
\fi\endgroup%
{\renewcommand{\dashlinestretch}{30}
\begin{picture}(2874,3837)(0,-10)
\put(117,2892){\arc{210}{1.5708}{3.1416}}
\put(117,3357){\arc{210}{3.1416}{4.7124}}
\put(582,3357){\arc{210}{4.7124}{6.2832}}
\put(582,2892){\arc{210}{0}{1.5708}}
\path(12,2892)(12,3357)
\path(117,3462)(582,3462)
\path(687,3357)(687,2892)
\path(582,2787)(117,2787)
\path(237,2787)(87,2562)
\path(462,2787)(612,2562)
\put(187,3087){$\ket{\ph_i}$}
\put(2292,2892){\arc{210}{1.5708}{3.1416}}
\put(2292,3357){\arc{210}{3.1416}{4.7124}}
\put(2757,3357){\arc{210}{4.7124}{6.2832}}
\put(2757,2892){\arc{210}{0}{1.5708}}
\path(2187,2892)(2187,3357)
\path(2292,3462)(2757,3462)
\path(2862,3357)(2862,2892)
\path(2757,2787)(2292,2787)
\path(2412,2787)(2262,2562)
\path(2637,2787)(2787,2562)
\put(2362,3087){$\ket{\ps_i}$}
\put(717,117){\arc{210}{1.5708}{3.1416}}
\put(717,582){\arc{210}{3.1416}{4.7124}}
\put(2232,582){\arc{210}{4.7124}{6.2832}}
\put(2232,117){\arc{210}{0}{1.5708}}
\path(612,117)(612,582)
\path(717,687)(2232,687)
\path(2337,582)(2337,117)
\path(2232,12)(717,12)
\path(687,912)(837,687)
\path(1212,912)(1062,687)
\path(1737,912)(1887,687)
\path(2262,912)(2112,687)
\thicklines
\path(387,2562)(912,1362)
\thinlines
\path(836.417,1459.914)(912.000,1362.000)(891.386,1483.963)
\thicklines
\path(2487,2487)(2037,1362)
\thinlines
\path(2053.713,1484.559)(2037.000,1362.000)(2109.421,1462.275)
\thicklines
\dashline{120.000}(687,3162)(2187,3162)
\put(787,1062){$\ket{\ph_i}$}
\put(1287,3312){$\cE_{AB}$}
\put(2412,1812){$R_B = 0$}
\put(-287,1812){$R_A = 1$}
\put(1117,312){$\ket{\ph_i}\ket{\ps_i}$}
\end{picture}
}

%% file: surprise.eepic
\setlength{\unitlength}{0.00077333in}
\begingroup\makeatletter\ifx\SetFigFont\undefined%
\gdef\SetFigFont#1#2#3#4#5{%
  \reset@font\fontsize{#1}{#2pt}%
  \fontfamily{#3}\fontseries{#4}\fontshape{#5}%
  \selectfont}%
\fi\endgroup%
{\renewcommand{\dashlinestretch}{30}
\begin{picture}(4817,5670)(0,-10)
\path(3150,3900)(3150,5025)
\path(3180.000,4905.000)(3150.000,5025.000)(3120.000,4905.000)
\path(3152,3885)(4277,4035)
\path(4162.018,3989.404)(4277.000,4035.000)(4154.088,4048.877)
\path(3150,3900)(4050,3300)
\path(3933.513,3341.603)(4050.000,3300.000)(3966.795,3391.526)
\dashline{60.000}(3150,3900)(3150,3075)
\dashline{60.000}(2775,3900)(4575,3900)
\dashline{60.000}(4050,3075)(2700,4350)
\dashline{60.000}(3150,1050)(3150,225)
\dashline{60.000}(2775,1050)(4575,1050)
\dashline{60.000}(4050,225)(2700,1500)
\path(3150,1050)(3150,2175)
\path(3180.000,2055.000)(3150.000,2175.000)(3120.000,2055.000)
\path(3152,1035)(4277,1185)
\path(4162.018,1139.404)(4277.000,1185.000)(4154.088,1198.877)
\path(3150,1050)(4275,1050)
\path(4155.000,1020.000)(4275.000,1050.000)(4155.000,1080.000)
\path(525,1050)(1725,1050)
\path(1605.000,1020.000)(1725.000,1050.000)(1605.000,1080.000)
\path(525,1050)(675,2175)
\path(688.877,2052.088)(675.000,2175.000)(629.404,2060.018)
\dashline{60.000}(525,2175)(525,750)
\dashline{60.000}(525,1050)(225,1050)
\path(450,3900)(1650,3900)
\path(1530.000,3870.000)(1650.000,3900.000)(1530.000,3930.000)
\path(450,3900)(600,5025)
\path(613.877,4902.088)(600.000,5025.000)(554.404,4910.018)

\path(420,4925)(450,5025)(480,4925)
\path(450,5025)(450,3900)

\path(495,2075)(525,2175)(555,2075)
\path(525,2175)(525,1050)

\dashline{60.000}(450,5025)(450,3600)
\dashline{60.000}(450,3900)(150,3900)
\put(4725,3825){$\ket{0}$}
\put(3075,5300){$\ket{\ps_2}=\ket{1}$}
\put(4275,4125){$\ket{\ps_1}$}
\put(4125,2850){$\ket{2}$}
\put(4050,3375){$\ket{\ps_3}$}
\put(4725,975){$\ket{0}$}
\put(3975,750){$\ket{\ps_3'}$}
\put(3075,2250){$\ket{\ps_2'}$}
\put(4125,0){$\ket{2}$}
\put(3675,1275){$\ket{\ps_1'}$}
\put(375,5150){$\ket{\ph_3}=\ket{1}$}
\put(0,2700){(b)}
\put(0,5475){(a)}
\put(1725,3825){$\ket{\ph_1}=\ket{0}$}
\put(750,4900){$\ket{\ph_2}$}
\put(425,2325){$\ket{\ph_3'}=\ket{1}$}
\put(825,1975){$\ket{\ph_2'}$}
\put(1800,1050){$\ket{\ph_1'}$}
\end{picture}
}